# Finite-Temperature Magnetism and Longitudinal Spin Transport in Compensated Half-Metal Ferrimagnets with Near-Zero Magnetization

Jiahao Shentu[1,a], Jie Li[1,a], Jianting Dong[1], Meng Zhu[2], Chen Wang[1], Qing Zhang[1], Yizhuo Song[1], Wenqi Sun[1], and Jia Zhang[1*]

[1]*School of Physics and Wuhan National High Magnetic Field Center, Huazhong University of Science and Technology, 430074 Wuhan, China.*

[2]*College of Physics and Electronic Science, Hubei Normal University, 435002 Huangshi, China.*

[a]These authors contributed equally to this work
[*]jiazhang@hust.edu.cn

## Abstract

Fully compensated ferrimagnets, characterized by nearly vanishing magnetic moments at zero temperature, perfect spin-polarized electronic structures, and ultrafast spin dynamics, represent a highly promising materials platform for next-generation spintronic devices. By minimizing stray magnetic fields and reducing switching currents, FCFs enable a viable path toward high-speed, low-power spintronic devices. In this work, we present a systematic investigation of the electronic structure, finite-temperature magnetism, and spin-dependent transport properties of the Heusler-type fully compensated half-metallic ferrimagnet $Mn_3Al$ in magnetic tunnel junctions. We demonstrate that $Mn_3Al$ is a half-metallic FCF with 100% spin polarization and fully compensated magnetization at 0 K. Combining density functional theory with Monte Carlo simulations, we characterize its finite-temperature magnetic properties and show that $Mn_3Al$ exhibits a net magnetization below 11.6 emu/cc at 300 K, confirming the robustness of its nearly compensated ferrimagnetic order at room temperature. Furthermore, we predict that the $Mn_3Al/MgO/Mn_3Al$ MTJ yields a giant tunneling magnetoresistance effect at 0 K, which remains as high as 232% at 300 K. Our results establish $Mn_3Al$ and related half-metallic FCFs as excellent candidates for magnetoresistance-based spintronic applications.

## Introduction

A key challenge in modern spintronics is the identification of magnetic materials that combine high spin polarization with low net magnetization—a property essential for next-generation nonvolatile memories featuring high integration density and ultralow power dissipation[1]. Traditional ferromagnets can generate spin-polarized currents, whereas detrimental stray fields are introduced and high switching currents are required, which severely hinder device miniaturization. Antiferromagnets, in contrast, exhibit no stray fields and allow ultrafast spin dynamics, yet their spin-degenerate electronic structures generally preclude efficient spin-polarized transport[2]. The recent discovery of altermagnets, which display nonrelativistic momentum-resolved spin-splitting and zero net magnetization protected by crystal symmetry, has opened a new avenue for zero-moment spintronic systems[3,4]. However, the number of confirmed metallic altermagnets remains very limited. $RuO_2$, a prototypical *d*-wave altermagnet, has been extensively studied theoretically, with sizable tunneling magnetoresistance (TMR) ratio predicted for antiferromagnetic tunnel junctions[5]. Despite this progress, the antiferromagnetic order in $RuO_2$ has not yet been conclusively established[6,7].

Fully compensated ferrimagnets (FCFs) offer a promising materials pathway to overcome the longstanding trade-off between spin polarization and stray field effects in spintronics. Initially proposed in the context of half-metallic antiferromagnets, the field of FCF research has advanced significantly through the discovery of several promising Heusler alloy systems[8]. These materials are distinguished by their ability to realize a fully compensated net magnetic moment at zero temperature while retaining half-metallic band structure, yielding nearly 100% spin polarization at the Fermi level and suppressing stray-field-induced perturbations[9]. Within this family, compounds with $D0_3$-type structure have been proposed via first-principles calculations[10]. $Mn_3Al$ serves as a representative example, exhibiting both high Curie temperature ($T_C \approx 605$ K) and a nearly perfect compensated magnetic order, making it one of the most attractive FCF candidates for room-temperature applications[11].

The potential of FCFs such as $Mn_3Al$ for advanced spintronic devices is substantial. Their half-metallic character enables its high spin-polarization in magnetic tunnel junctions (MTJs), which can in principle lead to remarkable TMR ratio[12]. Compared with altermagnets, which possess strictly zero net magnetization and often impose severe constraints on magnetic state manipulation[13,14], FCFs offer unique practical advantages: the small, temperature-sensitive net magnetization allows efficient control via external magnetic fields, spin-transfer torques[15,16] and spin-orbit torques[17,18] Moreover, many FCFs inherit high Curie temperatures from their Heusler parent compounds. A key parameter for reliable device operation that remains an active area of optimization in many emerging altermagnet systems.

In this work, we present a comprehensive first-principles study linking the fundamental properties of the prototypical Heusler-based FCF $Mn_3Al$ to its performance in MTJ, along with a realistic assessment of its finite-temperature magnetic properties. We first confirm its half-metallic ground state and compensated collinear ferrimagnetic order. We then calculate magnetic exchange interactions and employ Monte Carlo (MC) simulations to predict the magnetization versus temperature (*M*-*T*) curve. Finite-temperature analysis reveals that robust magnetic compensation persists at low temperatures, with a small net moment emerging upon heating—orders of magnitude smaller than that of conventional ferromagnets. Finally, we demonstrate that a $Mn_3Al$/MgO/$Mn_3Al$ MTJ, leveraging perfect spin polarization, achieves a theoretically infinite TMR at zero temperature, and retains spin-polarization over 70% at room temperature. These results highlight the potential of $Mn_3Al$ for applications with low switching current requirements, minimal stray fields, and high thermal stability.

## Computational method

First-principles calculations of bulk $Mn_3Al$ were performed within density functional theory (DFT) using the VASP package[19-21]. Electron-ion interactions were described by the projector augmented wave (PAW) method[22], and the exchange-correlation potential[23] was treated within the generalized gradient approximation (GGA) using the Perdew-Burke-Ernzerhof (PBE) parametrization. The calculations employed the primitive face-centered cubic (fcc) primitive cell, with a plane-wave cutoff energy of 600 eV and an $18\times18\times18$ ***k***-point mesh.

To characterize the finite-temperature magnetism of $Mn_3Al$, the Heisenberg exchange interaction parameters were calculated using the VASP and TB2J packages[24,25]. Maximally localized Wannier functions for all $Mn_3Al$ orbitals were constructed with Wannier90[26] to generate a tight-binding Hamiltonian, from which isotropic exchange parameters were extracted via the Green's function method implemented in the TB2J package. A 30×30×30 ***k***-point mesh ensured convergence of the exchange parameters. The Curie temperature and *M-T* curve of $Mn_3Al$ were computed using the Uppsala Atomistic Spin Dynamics (UppASD) code[27], on the basis of the Landau-Lifshitz-Gilbert (LLG) equation and MC simulations with a 20×20×20 supercell.

TMR for $Mn_3Al$-based MTJs was calculated using the Quantum ESPRESSO package[28]. The in-plane lattice constant of the $Mn_3Al/MgO/Mn_3Al$ MTJ was fixed to the theoretical equilibrium value of bulk $Mn_3Al$ (4.101 Å), imposing a 2.68% tensile strain on the MgO barrier. The scattering region of the MTJ stack consists of two $Mn_3Al$ unit cells for each electrode and one MgO unit cell as the tunneling barrier. Zero-bias spin-resolved ballistic transport was computed within the Landauer-Büttiker formalism using the PWCOND module. Transmission spectra $T(E,\boldsymbol{k}_{\|})$ at the Fermi level were evaluated on a 150×150 $\boldsymbol{k}_{\|}$ grid. The conductance for spin channel $\sigma$ reads:

$$G_{\sigma}=\frac{e^2}{h}\sum_{\boldsymbol{k}_{\|}}T_{\sigma}\left(\boldsymbol{k}_{\|}\right) \tag{1}$$

where $T_{\sigma}(\boldsymbol{k}_{\|})$ is the transmission probability for an electron at the Fermi energy with spin $\sigma$ and Bloch wave vector $\boldsymbol{k}_{\|} = (k_x, k_y)$, $e$ is the elementary charge, and $h$ is Plank's constant. The TMR ratio is defined as TMR = $(G_P - G_{AP})/G_{AP}$, with $G_P$ and $G_{AP}$ are the conductances for parallel and antiparallel magnetic alignments, respectively.

Calculations of the Bloch spectral function (BSF), spin conductivity, and spin polarization of $Mn_3Al$ at room temperature (300 K) were performed using the spin-polarized relativistic Korringa-Kohn-Rostoker (SPR-KKR) package[29]. Self-consistent calculations employed 32 energy points, a 45×45×45 ***k***-point mesh in the Brillouin zone, and an angular momentum cutoff $l_{max}$ = 3 for the wave-function expansion. For the BSF, a 52×52×52 ***k***-point mesh was used. To ensure convergence of the transport properties, a dense 782×782×782 ***k***-point mesh was employed. The conductivity and spin conductivity tensors were computed based on the Kubo-Bastin linear response formalism[30]. Phonon scatterings at finite temperatures was accounted for using the alloy analogy model within the Coherent Potential Approximation (CPA)[31-34]. To incorporate lattice vibrations explicitly, 14 phonon displacement directions (NVIBRA) are included. From the resulting conductivity data, the spin polarization of $Mn_3Al$ at 300 K was estimated.

## Results and discussions

### Half-Metallic Electronic Structure and Finite-Temperature Magnetism of $Mn_3Al$

As illustrated in Fig. 1(a), $Mn_3Al$ crystallizes in the cubic $D0_3$-type structure (space group $Fm\bar{3}m$, No. 225), [9-11]. Its magnetic ground state is a FCF, with two crystallographically inequivalent Mn sublattices (Mn(A) and Mn(B)) antiparallelly aligned. Our first-principles calculations, performed using the primitive fcc cell (containing two Mn(A), one Mn(B), and one Al atom), yield atomic magnetic moments of 1.418, −2.820, and −0.023 $\mu_B$ for Mn(A), Mn(B), and Al, respectively. This confirms the opposite alignment of the Mn(A) and Mn(B) moments. As expected from the stoichiometry and crystal structure, the magnetic moment on the Mn(B) is about twice of that on Mn(A). Given that Mn(A) atoms are twice as numerous as Mn(B), a nearly perfect magnetic compensation is achieved at 0 K, consistent with prior work[10].

The spin-resolved band structure [Fig. 1(c)] exhibits a distinct half-metallic character: the spin-up channel displays a band gap of 0.49 eV at the Γ point, while the spin-down channel is metallic and crosses the Fermi level. This yields 100% spin polarization at $E_F$, which constitutes the fundamental prerequisite for high-performance spintronic applications. Further insight is provided by the atom-projected density of states (PDOS) shown in Figs. 1(d)-1(g). The states near $E_F$ are dominated by the *d* orbitals of Mn(A) and Mn(B), which exhibit strong and opposite spin polarization, consistent with their antiferromagnetic coupling. In contrast, Al atoms display negligible spin polarization and magnetic moment. Importantly, within the energy window from −0.2 to 0.2 eV around $E_F$, the PDOS for all atomic species consists entirely of spin-down states. This unequivocally confirms the robust half-metallicity and accounts for the fully compensated magnetic state of $Mn_3Al$[2,9,35,36].

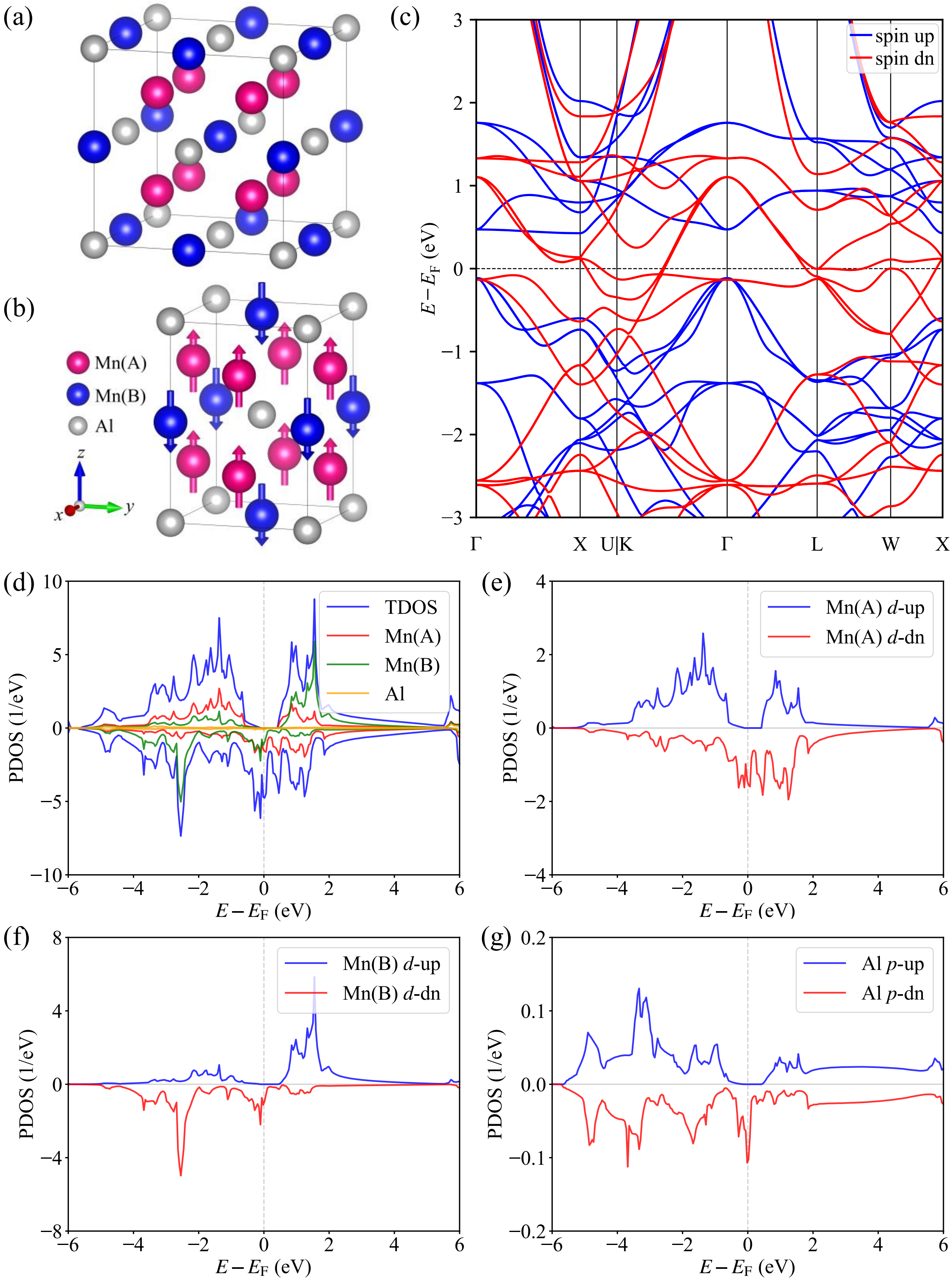

FIG. 1. (a) Crystal structure of $Mn_3Al$ in the conventional cubic unit cell. (b) Magnetic unit cell with magnetic moments (denoted by arrows) aligned along the *z* axis. (c) Spin-up (blue curves) and spin-down (red curves) band structures of bulk $Mn_{3S}Al$. (d-g) Spin-polarized PDOS for each distinct atomic site of $Mn_3Al$, calculated for the primitive fcc cell with cubic lattice constants a = 5.800 Å.

Shown are the total DOS, Mn(A)-*d* DOS, Mn(B)-*d* DOS, and Al-*p* DOS. The Fermi energy is set to 0 eV.

To uncover the microscopic origin of finite-temperature magnetism in $Mn_3Al$, we determined the Heisenberg exchange parameters using the VASP + TB2J approach. The fitting of VASP band structure to Wannier functions is detailed in Supplementary Note 1. Neglecting spin-orbit coupling, the spin Hamiltonian is expressed as:

$$\mathcal{H} = -\sum_{i \neq j} J_{ij} \boldsymbol{e}_i \cdot \boldsymbol{e}_j \tag{2}$$

where $\boldsymbol{e}_i$ and $\boldsymbol{e}_j$ denote unit spin vectors and $J_{ij}$ is the isotropic exchange energy between atoms i and j. The variation of $J_{ij}$ with the Mn-Mn interatomic distance is shown in Fig. 2(a). The nearest-neighbor Mn(A)-Mn(B) interaction is strongly antiferromagnetic ($J_1 = -10.8$ meV) and constitutes the dominant magnetic coupling. By contrast, the next-nearest-neighbor Mn(A)-Mn(A) exchange is ferromagnetic ($J_2 = 7.9$ meV). Together, these couplings stabilize a ground state with ferromagnetic alignment within the Mn(A) sublattice and antiparallel alignment between Mn(A) and Mn(B) moments. Although the third-neighbor Mn(B)-Mn(B) exchange is antiferromagnetic ($J_3 = -2.5$ meV), the much stronger $J_1$ interaction dictates the moment orientation of Mn(B). Because Al carries a negligible magnetic moment, all exchange couplings involving Al are omitted. Notably, as illustrated in Fig. 3(b), the $J_5$ exhibits anisotropic coupling within Mn(A)-Mn(A) pairs at 5.023 Å with $J_5^1 = 0.76$ meV mediated by an intervening Mn(B) atom, and $J_5^2 = -0.19$ meV mediated by an intervening Al atom. A further splitting is observed at 8.700 Å, yielding three distinct exchange parameters: 0.09, 0.01, and −0.05 meV.

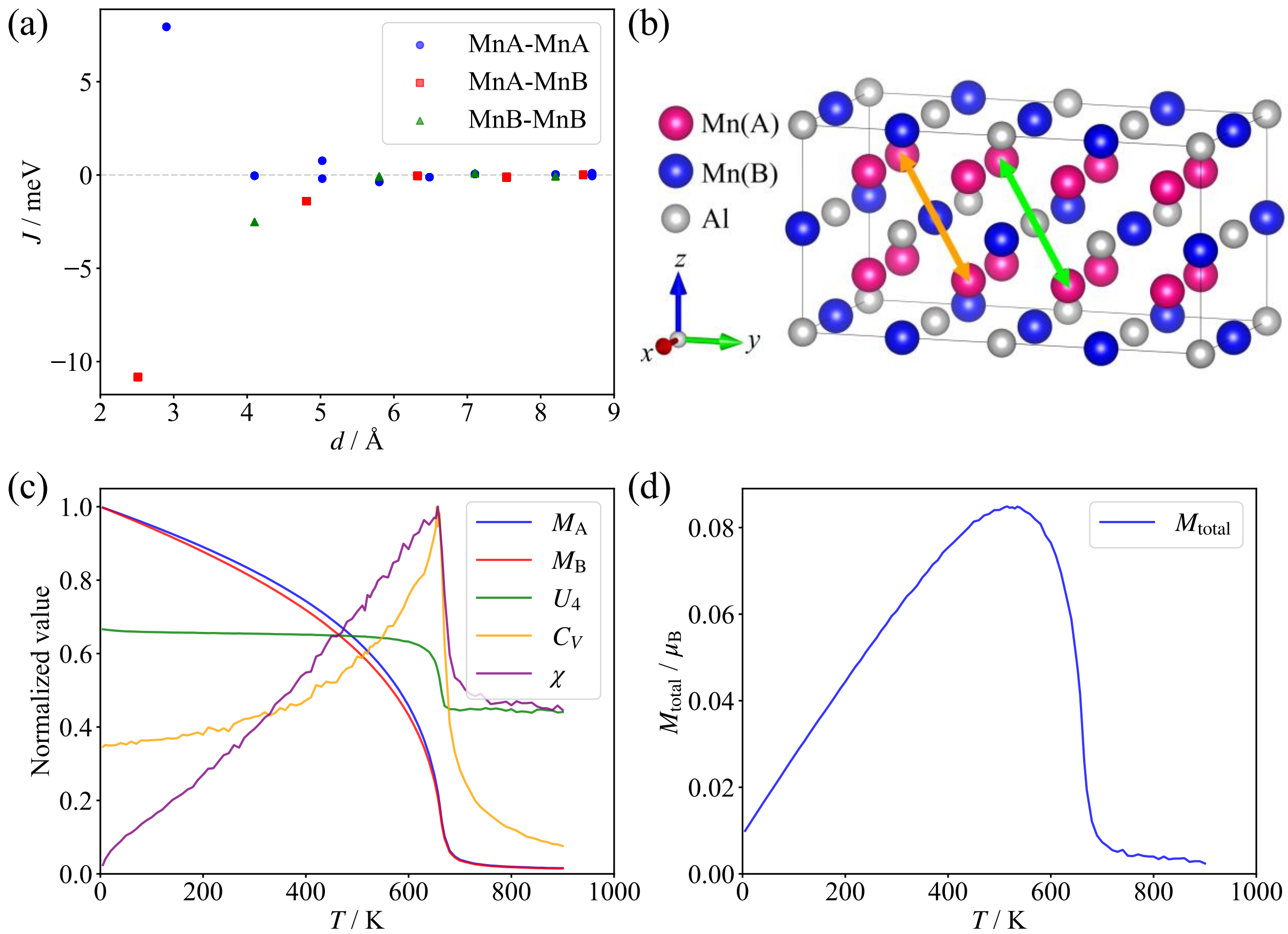

FIG. 2. (a) Exchange coupling parameter $J_{ij}$ as a function of Mn-Mn interatomic distance. Different symbols correspond to distinct Mn-Mn atomic pairs. (b) Schematic illustration of an Mn(A) atom pair separated by 5.023 Å, where the midpoint site can be occupied by either Mn(B) or Al. (c) Temperature dependence of the sublattice magnetic moments of Mn(A) and Mn(B), the specific heat, the magnetic susceptibility, and the Binder cumulant $U_{Binder}$. (d) Total magnetic moment of $Mn_3Al$ obtained from Monte Carlo simulations.

The finite-temperature magnetism of $Mn_3Al$ was investigated by MC simulations, with results shown in Fig. 2(c). The sublattice magnetizations $M_{MnA}$ and $M_{MnB}$ are normalized to their zero-temperature values. Thermodynamic signatures of the magnetic phase transition include the constant-volume heat capacity $C_V$, the isothermal magnetic susceptibility $\chi$, and the fourth-order Binder cumulant $U_4 = 1 - \frac{\langle M^4 \rangle}{3\langle M^2 \rangle^2}$ [37], where $M$ in $U_4$ denotes the net magnetic moment of the system. To determine the Curie temperature $T_C$ accurately from finite-size supercell, we employed Binder's cumulant-crossing method: for a sufficiently large system, $U_4$ approaches 2/3 below $T_C$ and 4/9 above it, consistent with our calculations. Combing $U_4$ with the other transition criteria, yields $T_C \approx 657$ K for $Mn_3Al$, in close agreement with the experimental value of 605 K[11].

The net magnetic moment per formula unit (considering only Mn(A) and Mn(B)) is plotted in Fig. 2(d). The total moment $M_{total}$ first increase, reaches a maximum of ~0.085 $\mu_B$ between 510 and 540 K, and then decrease to zero at $T_C$. At 300 K, $M_{total} = 0.061$ $\mu_B$, consistent with the experimental value of 0.11 ± 0.04 $\mu_B$[11]. As listed in Table 1, the net magnetization of $Mn_3Al$ is 11.6 emu/cc, which is roughly two orders of magnitude smaller than that of the conventional ferromagnets CoFeB used in MgO-MTJ[38], comparable to that of altermagnetic CrSb thin film[39], and only several times larger than those of the canted antiferromagnet $\alpha$-$Fe_2O_3$[40] and the noncollinear AFMs $Mn_3Sn$ (0.67 emu/cc)[41,42] and $Mn_3Pt$[43]. This confirms that $Mn_3Al$ retains an almost fully compensated magnetic state with a small net magnetization even at finite temperatures. In a MTJ, the spin-transfer torques switching current density is linearly proportional to the magnetization of the free layer[44]. Therefore, the compensated ferrimagnet $Mn_3Al$ should be promising for high-density and low-power-consumption MTJ-based device.

Table 1. Net magnetization of $Mn_3Al$ at 300 K, compared with antiferromagnets with weak ferromagnetism and ferromagnet CoFeB used in MgO-MTJ.

| Materials | Net magnetization (emu/cc) |
|---|---|
| Noncollinear $Mn_3Sn$ [41,42] | <1 |
| Noncollinear $Mn_3Pt$[43] | ~2 |
| Canted AFM $\alpha$-$Fe_2O_3$ [40] | ~3 |
| Altermagnetic CrSb thin film[39] | ~10 |
| $Mn_3Al$ (This work) | 11.6 |
| CoFeB [38] | ~1300 |

By taking SOC effect, $Mn_3Al$ exhibits cubic magnetocrystalline anisotropy, which can be described by the expression:

$$\mathcal{H}_{\text{ani}} = K_1\left(m_x^2 m_y^2 + m_y^2 m_z^2 + m_z^2 m_x^2\right) + K_2 m_x^2 m_y^2 m_z^2 \quad (3)$$

where $K_1$, $K_2$ are the cubic anisotropy energy constants and $\boldsymbol{m} = (m_x, m_y, m_z)$ is the unit vector along the magnetic moment direction. The magnetocrystalline anisotropy energy was calculated by treating SOC as a second-order perturbation within the force theorem [45]. Band energies were evaluated with the magnetic moments aligned along the [001], [011], and [111] directions of the cubic cell. Using the primitive cell, we obtain energy differences of $E_{[011]} - E_{[001]}$ = 3.5 μeV and $E_{[111]} - E_{[001]}$ = 3.2 μeV. These results identify the [001] direction as the magnetic easy axis of $Mn_3Al$. The fitted cubic anisotropy constants are $K_1$ = 14.1 μeV, $K_2$ = −39.8 μeV. The corresponding three-dimensional anisotropy energy surface is shown in Fig. 3.

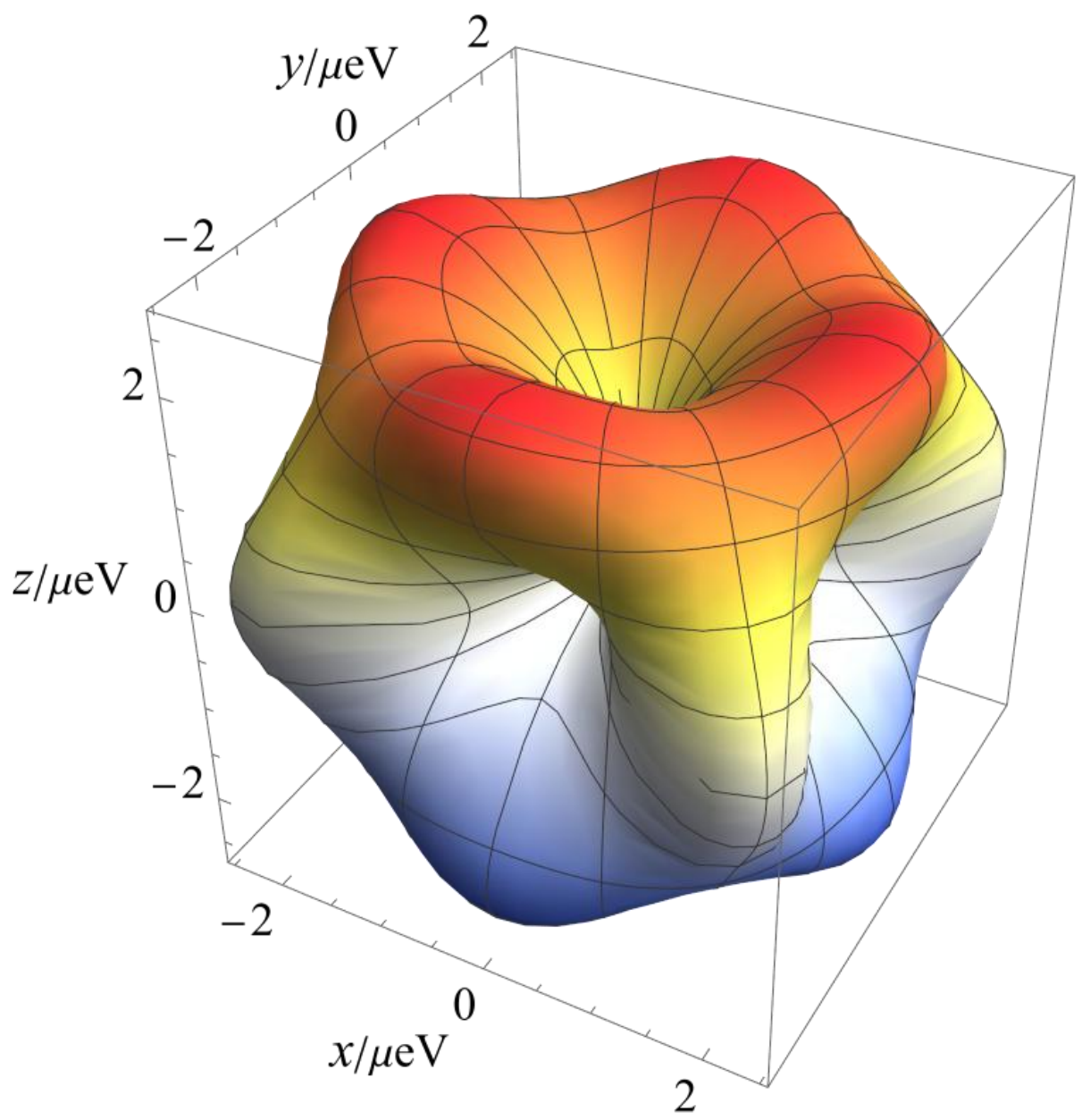

FIG. 3. Calculated magnetocrystalline anisotropy energy surface of $Mn_3Al$. The radial distance from the origin represents the magnitude of the anisotropy energy. The colored surface for better demonstration.

## Tunneling magnetoresistance in $Mn_3Al$ MTJ and spin polarization at finite temperature

The half-metallic character of bulk $Mn_3Al$ is essential for achieving high TMR in MTJ, as unequivocally demonstrated by its spin-resolved Fermi surface. As shown in Fig. 4(a), the spin-down channel exhibits a complex, multi-connected Fermi surface that spans a large fraction of the two-dimensional Brillouin Zone (2D BZ), confirming the presence of abundant Bloch states at $E_F$. The $\boldsymbol{k}_{\parallel}$-resolved ballistic transmission of bulk $Mn_3Al$ across the 2D BZ, presented in Fig. 4(b), provides direct insight into its effectiveness as an MTJ electrode. The 100% spin polarization at $E_F$ guarantees that, within a coherent tunneling picture where spin and in-plane crystal momentum ($\boldsymbol{k}_{\parallel}$) are conserved, the conductance of a parallel (P) aligned $Mn_3Al/MgO/Mn_3Al$ junction is finite and dominated by spin-down electrons. In contrast, for the antiparallel (AP) configuration, spin-down electrons injected from one electrode encounter a band gap in the same spin channel of the opposing electrode, leading to complete suppression of tunneling. This yields an infinite TMR at 0 K, originating from the vanishing conductance in the AP state.

To quantitatively validate this prediction, we constructed a realistic $Mn_3Al/MgO/Mn_3Al$ MTJ. We first examined six possible interfacial stacking configurations [Fig. 4(d)] and identified the Ba-stacked structure as the energetically most favorable. Subsequent optimization of the interfacial separation for this configuration yielded an equilibrium $Mn_3Al$-MgO distance of 2.07 Å, which was employed in all subsequent transport calculations. The calculated $\boldsymbol{k}_{\parallel}$-resolved transmission of the MTJ with the optimized interface confirms our predictions. In the P configuration [Fig. 4(e)], the transmission pattern closely mirrors that of the bulk $Mn_3Al$ electrodes: transmission is fully suppressed in the spin-up channel and remains finite in the spin-down channel [Fig. 4(c)]. The MgO barrier acts as a momentum filter, suppressing transmission at specific points—particularly near the zone boundary—while preserving the fundamental spin-channel imbalance. The situation is drastically different for the AP configuration [Fig. 4(f)]: spin-down (majority) carriers from one electrode become spin-up (minority) carriers in the opposing electrode, where they encounter a band gap. The half-metallic gap thus completely suppresses electron tunneling in the AP state.

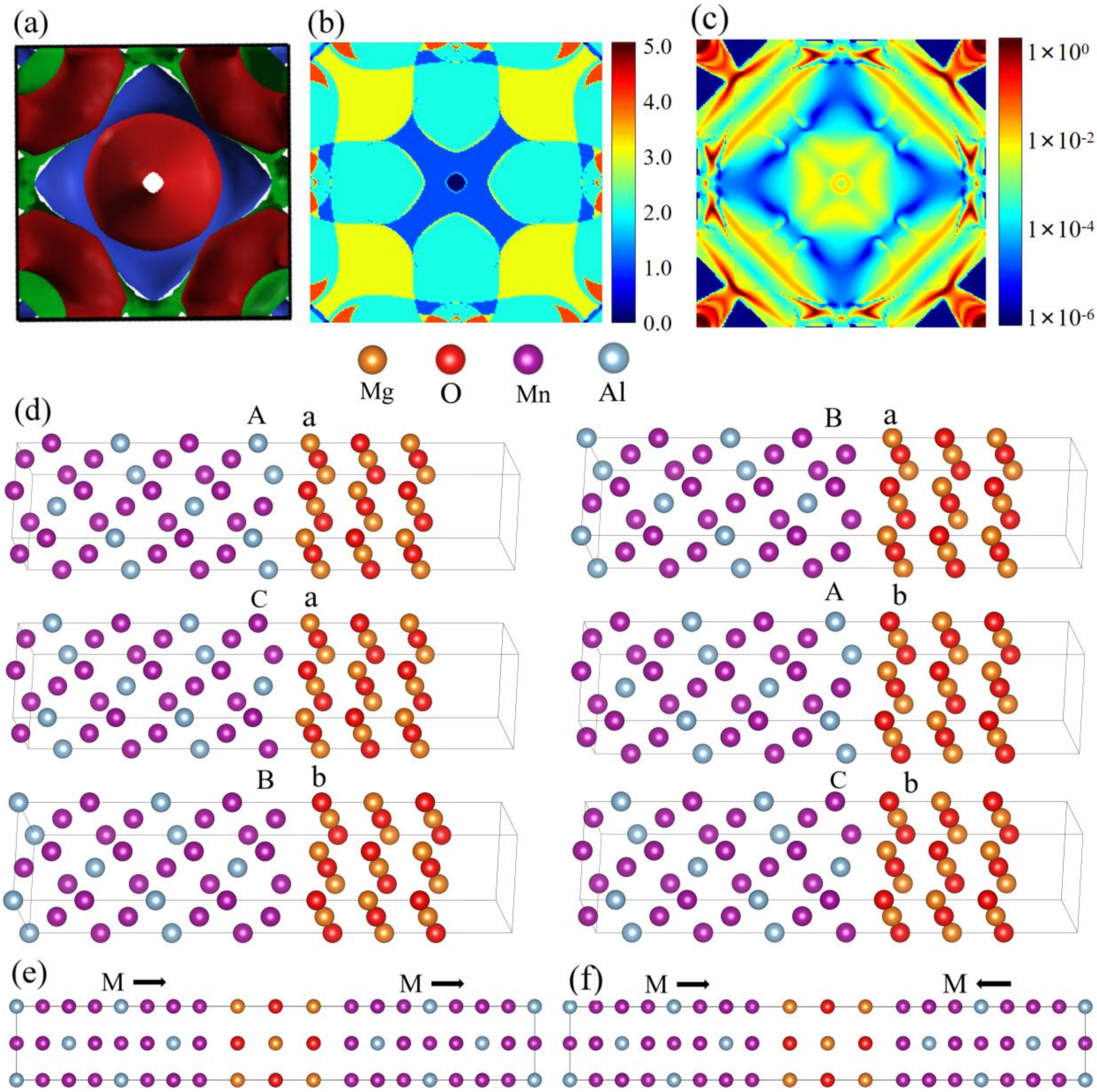

FIG. 4. (a) Spin-down channel Fermi surface of bulk $Mn_3Al$. (b) Ballistic transmission in the spin-down channel for bulk $Mn_3Al$; the color bar indicates the number of conducting channels in the 2D BZ. (c) Spin-down-resolved electron transmission distribution in the 2D BZ for the $Mn_3Al$/MgO/$Mn_3Al$ in the parallel magnetization configuration. (d) Side view of $Mn_3Al$ (2 unit cells, u.c.)/MgO (1 u.c.)/vacuum structures, showing the six possible interfacial stacking configurations between $Mn_3Al$ and MgO. (e), (f) Side views of the $Mn_3Al$/MgO/$Mn_3Al$ in the (e) parallel and (f) antiparallel magnetization configurations. Arrows denote the magnetic moment directions of the Mn atoms.

Integrating the transmission over the 2D BZ yields the total spin-resolved transmission of the $Mn_3Al$/MgO/$Mn_3Al$ junction at $E_F$: $T_{P\uparrow} = 0$, $T_{P\downarrow} = 0.023$ for parallel magnetization state, and $T_{AP\uparrow} = T_{AP\downarrow} = 0$ for the antiparallel state. Using the standard definition of the TMR ratio, TMR $= (G_P - G_{AP}) / G_{AP}$, with the conductance $G \propto T$, the vanishing conductance in the AP configuration results in a theoretically infinite TMR outperforms of noncollinear antiferromagnetic TMR[46], altermagnetic TMR[47,48] and van der Waals TMR[49]. This finding highlights the exceptional potential of $Mn_3Al$ for ultrahigh-density nonvolatile memory devices.

However, at finite temperatures, spin-flips scatterings induced by phonons and spin disorder can degrade the half-metallic character (100% spin-polarization) and lead to a finite tunneling magnetoresistance in MTJs. To evaluate the spin transport properties of $Mn_3Al$ at finite temperatures, we calculated the BSF, spin conductivity and subsequently evaluate the spin polarization using the SPRKKR package. The effects of phonon scattering were taken into account, and the fcc primitive cell was employed. The Cartesian *x*, *y*, *z* axes correspond to the *a*, *b*, *c* axes of the cubic unit cell, respectively. The magnetic moments of Mn(A) atoms were oriented along the negative *z*-axis, while those of Mn(B) atoms were oriented along the positive *z*-axis.

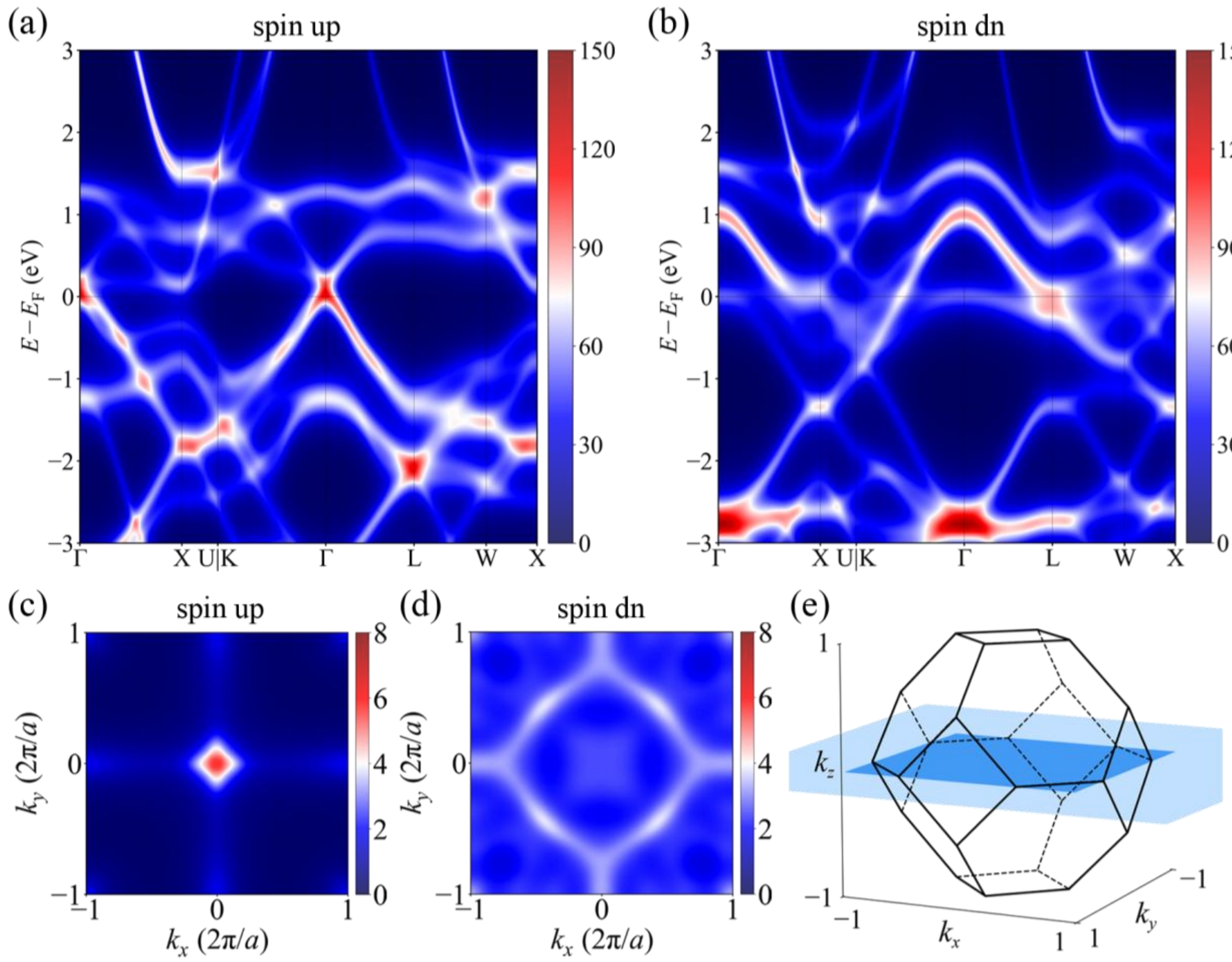

FIG. 5. BSF of $Mn_3Al$. (a) and (b) BSF along high-symmetry ***k*** paths in the Brillouin zone. (c) and (d) BSF at the Fermi level on a horizontal cut at $k_z = 0$, shown in the $(k_x,k_y)$ plane with coordinates in units of $2\pi/a$ ($a$ = 5.8 Å). (a) and (c) correspond to the spin-up projection, (b) and (d) to the spin-down projection. (e) Brillouin zone indicating the location of the selected horizontal cross section (dark blue).

Fig.5 shows the spin-projected BSF of $Mn_3Al$ at room temperature (300 K), calculated with phonon scattering included via the CPA. Panels (a) and (b) present the BSF along high-symmetry ***k***-point paths in the Brillouin zone, while panels (c) and (d) display a cross section at the Fermi energy and $k_z = 0$. Compared with the 0 K band structure (FIG. 1(c)), the bands at 300 K are broadened by phonon scattering, reflecting a finite electronic lifetime and yielding a more diffuse and complex distribution in the Brillouin zone. Notably, the spin-up band at 0 K exhibits a band gap of 0.49 eV at the Γ point, which is closed at 300 K by thermal broadening. Consequently, a significant electron population appears near the Fermi level around the Γ point, as is particularly evident in panel (c). The modifications of the electron distribution within the Brillouin zone, especially near the Fermi level, have a crucial impact on the transport properties of $Mn_3Al$.

Furthermore, we calculated the charge and spin conductivities of $Mn_3Al$. The longitudinal charge conductivity (unit: S/cm) $\sigma_{xx} = \sigma_{yy} = 6913$ , $\sigma_{zz} = 7379$ and the anomalous Hall conductivity $\sigma_{xy} = 264$. These anisotropic features originate from spin-orbit coupling (SOC). The *z*-component longitudinal spin conductivity (unit: ℏ/2e S/cm), $\sigma_{xx}^{z} = \sigma_{yy}^{z} = 5837$ , $\sigma_{zz}^{z} = 5408$. Using the *z*-component of longitudinal charge conductivity and spin conductivity, the spin polarization of $Mn_3Al$ at 300 K is evaluated to be $P = \sigma_{zz}^{z} / \sigma_{zz} = 73.3\%$ . This indicates that $Mn_3Al$ largely retains its half-metallic character even at room temperature. For an $Mn_3Al$/MgO/$Mn_3Al$ MTJ, the TMR ratio is roughly estimated via Jullière's model as[50]: $\mathrm{TMR} = 2P^2 / (1-P^2) = 232\%$ . An alternative estimation of the spin polarization is obtained by calculating the DOS at 300 K including phonon scattering. The polarization $P$ is then determined from the DOS of the two spin channels at the Fermi level. Relevant details are provided in Supplementary Note 2. The resulting high TMR ratio demonstrates significant potential for applications in magnetic random-access memory (MRAM).

## Conclusion

In summary, we have performed a theoretical investigation of the compensated ferrimagnet $Mn_3Al$, identifying it as a high-performance material for next-generation spintronics. First-principles calculations confirm its half-metallic character and perfect magnetic compensation at 0 K. Our finite-temperature analysis, which integrates first-principles-derived exchange parameters with MC simulations, provides a realistic assessment of its finite-temperature behavior. We find that $Mn_3Al$ retains a net magnetic moment as small as 0.061 $\mu_B$ at room temperature, a value two orders of magnitude smaller than that of conventional ferromagnets such as bcc Fe. The resulting $Mn_3Al$/MgO/$Mn_3Al$ MTJ exploits this ideal electronic structure to deliver perfect spin-filtering and a theoretically infinite TMR ratio at 0 K. Moreover, we estimate that its spin polarization at 300 K remain as high as 73.3%. This unique combination of high spin polarization, low net magnetization, and high Curie temperature makes $Mn_3Al$ a compelling candidate for high-density, low-power spintronic devices such as MRAMs. It thus offers a versatile alternative that combines the advantages of traditional ferromagnets with those of emerging compensated magnets exhibiting nearly zero net magnetization.

## Supplementary Material

The supplementary material presents a comparison of the band structures calculated via VASP and Wannier90, the density of states of $Mn_3Al$ at 300 K accounting for phonon scattering, and our corresponding estimates of the spin polarization ratio and TMR derived from these results.

## Acknowledgement

This work was supported by the National Natural Science Foundation of China (grant No. T2394475, T2394470), and the National Key Research and Development Program of China (2024YFA1611200).